\documentclass[
    ,final            
  ]
  {aipproc}

\layoutstyle{6x9}

\usepackage{graphicx}

\providecommand{\tabularnewline}{\\}

\usepackage{psfrag}

\begin{document}

\title{Some issues about neutrino processes in color superconducting quark
matter}



\author{Qun Wang}{address={Department of Modern Physics, 
University of Science and Technology of China, Anhui 230026, China}}

\begin{abstract}
Several relevant issues in computing neutrino emissivity in Urca processes
in color superconducting quark matter are addressed. These include:
(1) The constraint on $u$ quark abundance is given from electric
neutrality and the triangle relation among Fermi momenta for participants.
(2) The phase space defined by Fermi momentum reduction of quarks
is discussed in QCD and NJL model. (3) Fermi effective model of weak
interaction is reviewed with special focus on its form in Nambu-Gorkov
basis. 
\end{abstract}

\maketitle

\section{Introduction}

Supernova explosions are among the most violent and spectacular phenomena
in our universe {[}for reviews, see, e.g. \cite{Bethe:1990mw,Burrows:1990ts}
and references therein, or see the talk by Sumiyoshi on this workshop
\cite{Sumiyoshi:2005ri}{]}. In later stage of the evolution of massive
stars, thermal-nuclear fusions which power the stars stop at the formation
of irons, the most stable nuclei. The collapse takes place when the
pressure cannot sustain gravitational forces due to drop of temperature.
Compact stars or sometime called neutron stars are one possible product
of such a collapse. Just before collapse the fraction of protons reach
the level of the most neutron-rich terrestrial matter, about 0.4,
much larger than that in nuclear matter in neutron stars. Therefore
the electron capture process $e+p\rightarrow n+\nu_{e}$ can occur
during the collapse because the Fermi surface of protons is high enough
to open up phase space. During the explosion the total energy released
in neutrino bursts can reach as much as 20\% of the solar mass. After
the explosion, the proton abundance falls to the lower level characteristic
of a compact star. The temperature of the newborn compact star exceeds
some tens of MeV \cite{Pethick:1991mk}. The compact star cools down
mainly by neutrino emissions in its earlier age and gamma-ray emissions
when it gets very old {[}for reviews of neutron star cooling, see,
e.g. \cite{Pethick:1991mk,Baym:1978jf,Yakovlev:2004iq}{]}. 

The baryon density in the core of a compact star is likely to reach
several times the nuclear saturation density, $\rho_{0}\sim0.16\;\mathrm{fm}^{-3}$
or $2.7\times10^{14}\mathrm{g}/\mathrm{cm}^{3}$. At such a high density,
nucleons in nuclear matter are crushed into their constituents, i.e.
quarks and gluons. This deconfinement transition to quark matter was
suggested by Collins and Perry already in 1975 \cite{Collins:1974ky}
based on the asymptotic freedom in quantum chromodynamics. In the
same paper they also mentioned the possibility that quark matter could
be a superfluid or a superconductor resulting from the attractive
inter-quark force in some channels. Barrois, Bailin and Love developed
this novel idea and studied the unusual variant of superconductivity
in quark matter, which we now call color superconductivity (CSC) \cite{Barrois:1977xd,Bailin:1983bm}.
They did their calculations in the framework of weak coupling approach
and did not take into account the dynamic screening of magnetic gluons
which are dominant agents in pairing quarks. Therefore the gap or
equivalently the transition temperature they obtained are too small
to be of relevance to any sizable observables. About fifteen years
later the color superconductivity had been re-discovered by several
groups who found the gap could be large enough to bring some real
effects in compact stars \cite{Alford:1997zt,Alford:1998mk,Rapp:1997zu,Son:1998uk,Pisarski:1999tv,Hong:1999fh}.
{[}For recent reviews on color superconductivity, see, for example,
\cite{Rajagopal:2000wf,Alford:2001dt,Reddy:2002ri,Casalbuoni:2003wh,Schafer:2003vz,Rischke:2003mt,Buballa:2003qv,Ren:2004nn,Shovkovy:2004me}.
Also see Zhuang's talk on this workshop based on 
Ref. \cite{He:2005bd,He:2005mp}.{]} 

\begin{figure}
\caption{\label{cap:main-neutrino}The main neutrino processes in quark matter
in normal and color superconducting states. }
\begin{minipage}[c]{1\textwidth}
\includegraphics[scale=0.6]{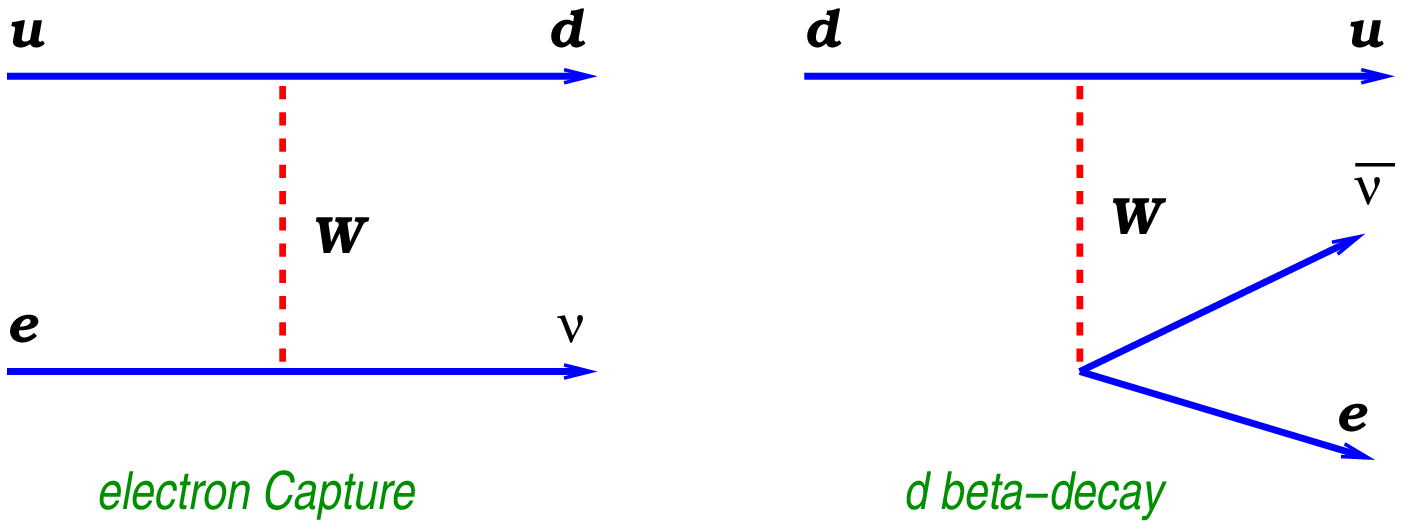} 
\hspace{0.2cm}
\includegraphics[scale=0.6]{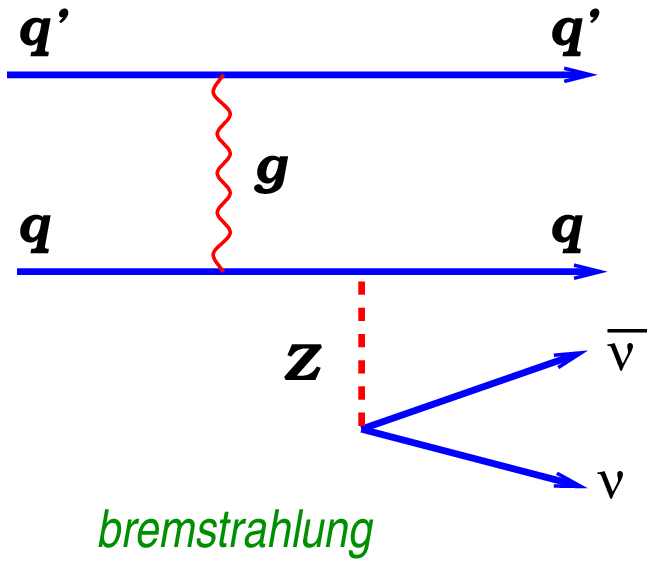}
\includegraphics[scale=0.6]{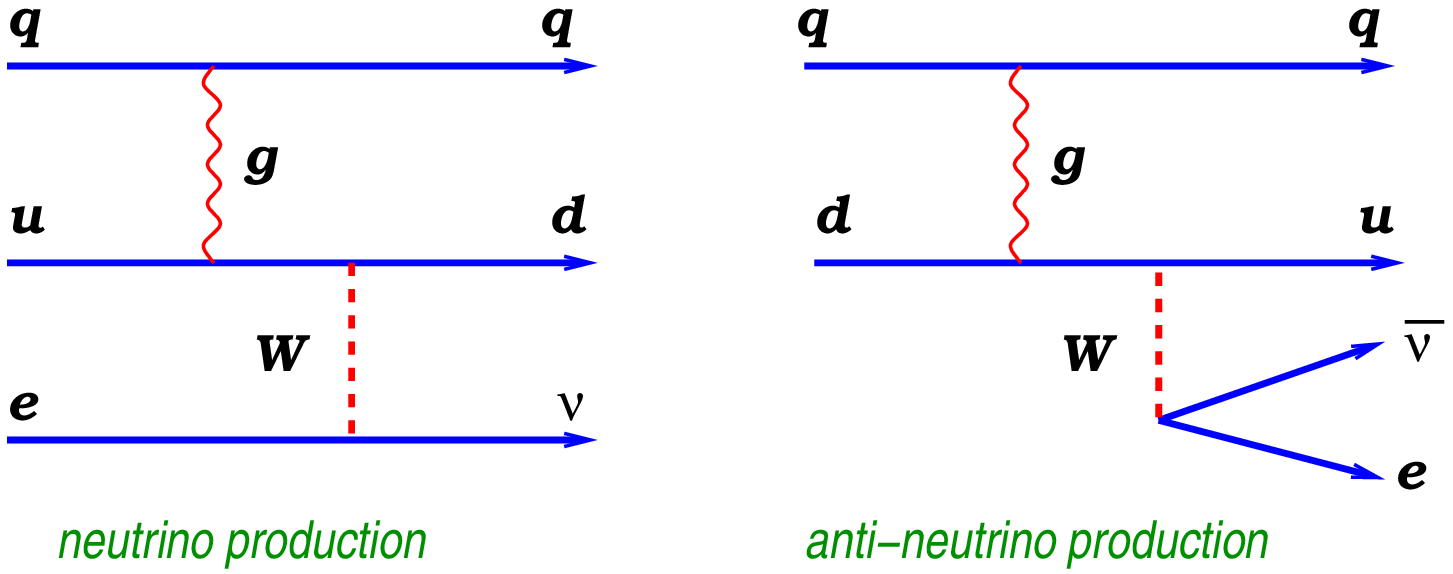}
\end{minipage} 
\end{figure}

Considering that neutrino emissions are main source of cooling for
young compact stars and that the star core is very likely to be CSC
quark matter, it is important to study neutrino emissions in a CSC
and how they influence the thermal history of the stars. It is well
known that the most efficient or fast cooling processes in quark matter
of normal state are the direct Urca processes, less efficient are
the modified Urca and bremsstrahlung processes or slow processes.
They are shown in Fig. \ref{cap:main-neutrino}. The direct Urca in
color superconducting quark matter was studied in Ref. \cite{Alford:2004zr,Schmitt:2005ee,Schmitt:2005wg,Jaikumar:2005hy,Wang:2006tg}.
In a CSC, besides these neutrino processes, there are other neutrino
sources such as decays of Goldstone modes in the color-flavor-locked
phase \cite{Reddy:2002xc,Jaikumar:2002vg} or in particular kaon condensed
phase of color flavor locked phase \cite{Reddy:2002ri}. In this paper,
we will briefly address several issues about direct Urca processes. 

\section{Constraint on $u$ quark abundance for direct Urca processes}

The name of Urca was coined by Gamow and Schoenberg when they studied
the star cooling mechanism by neutrino emissions \cite{Gamow1941}.
As Gamow commemorated, the name actually came from a Casino in Rio
de Janeiro considering that Urca processes' accounting for rapid energy
loss in stars is just like the casino's exhausting money from gamblers'
pockets \cite{Pethick:1991mk}. Later they gave Urca a physical meaning,
i.e. the abbreviation of \emph{unrecordable cooling agent}. 

In nuclear matter, the proton abundance is crucial to phase space
for Urca processes to proceed. In normal state quark matter, the abundance
of $u$ quarks plays an equal role. Suppose there are only light quarks
and electrons in the system and they are ultra-relativistic at high
baryon densities. Let us estimate the fraction of $u$ quarks from
the electric charge neutrality and $\beta$ equilibrium condition,
\begin{eqnarray}
\frac{2}{3}n_{u} & = & \frac{1}{3}n_{d}+n_{e},\;\;\;\;(\mathrm{electric\; neutrality})\nonumber \\
\mu_{d} & = & \mu_{u}+\mu_{e},\;\;\;\;(\beta\;\mathrm{equilibrium})\nonumber \\
p_{Fd} & < & p_{Fu}+p_{Fe}.\;\;\;\;(\mathrm{triangle\; relation})\label{eq:phase-neutrality}\end{eqnarray}
Here $n_{i}$, $\mu_{i}$ and $p_{Fi}$ are the number density, chemical
potential and Fermi momentum for particle $i$ respectively. The $\beta$
equilibrium condition means that it costs no energy to convert a $u$
quark and an electron to a $d$ quark and vice versa. In $\beta$
equilibrium the abundances of quarks and electrons are stable and
do not change with time. The chemical potential is just the Fermi
energy, so in $\beta$ equilibrium and at zero temperature there is
no phase space for Urca processes because participating particles
on their respective Fermi surfaces do not satisfy energy conservation.
At non-zero temperatures, quarks and electrons can be excited above
their Fermi surfaces of order $T$, energy conservation for Urca processes
can be reached in a small range. The third condition of Eq. (\ref{eq:phase-neutrality})
is the triangle relation among Fermi momenta of light quarks and electrons.
We know that at low temperature momenta of participants are close
to their Fermi surfaces. So in order to satisfy momentum conservation
$\mathbf{p}_{d}=\mathbf{p}_{u}+\mathbf{p}_{e}$(the neutrino momentum
is negligible), three momenta form a triangle in a plane, so the length
of each side should be be less than the sum of lengths of other twos. 

Rewriting the electric neutrality condition as $2p_{Fu}^{3}=p_{Fe}^{3}+p_{Fd}^{3}$
where the number densities are given by $n_{i}=p_{Fi}^{3}/\pi^{2}$
for quarks and $n_{e}=p_{Fe}^{3}/(3\pi^{2})$ for electrons. Applying
the triangle relation, we have \begin{eqnarray*}
2p_{Fu}^{3} & > & (p_{Fd}-p_{Fu})^{3}+p_{Fd}^{3}.\end{eqnarray*}
In terms of the ratio $p_{Fu}/p_{Fd}$, the above inequality becomes
\begin{eqnarray*}
3(p_{Fu}/p_{Fd})^{3}-3(p_{Fu}/p_{Fd})^{2}+3(p_{Fu}/p_{Fd})-2 & > & 0,\end{eqnarray*}
which leads to \begin{equation}
p_{Fu}/p_{Fd}>0.8.\label{eq:pfupfd}\end{equation}
From the $\beta$ equilibrium condition in (\ref{eq:phase-neutrality})
and the inequality (\ref{eq:pfupfd}), we obtain\begin{equation}
p_{Fe}/p_{Fd}<0.2.\label{eq:pfepfd}\end{equation}
The above inequality means the Fermi momentum of electrons must be
less than 20\% of that of $d$ quarks. Following (\ref{eq:pfupfd}),
the fraction of $u$ quarks then satisfies the inequality \begin{equation}
x=p_{Fu}^{3}/(p_{Fu}^{3}+p_{Fd}^{3})>0.8^{3}/(0.8^{3}+1)\approx1/3.\label{eq:u-fraction}\end{equation}
 Hence the fraction of $u$ quarks must exceed $1/3$ for Urca processes
to proceed. The factor $1/3$ comes naturally if the electron density
is neglected.

\section{Phase space for direct Urca processes}

\begin{figure}
\caption{\label{cap:phasespace}Phase space for DU processes with massless
quarks}
\includegraphics[scale=0.8]{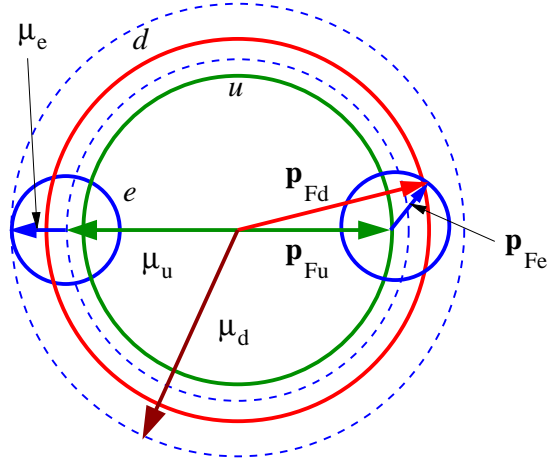}
\end{figure}

In ultra-relativistic case, the $\beta$ equilibrium condition is
not compatible with the triangle relation in (\ref{eq:phase-neutrality})
since the former requires that $p_{Fd}=p_{Fu}+p_{Fe}$. This means
phase space for neutrino emissions is zero for direct Urca. If the
quark-quark interaction is switched on, Fermi momenta are not equal
to chemical potentials any more, instead they get negative corrections
from Landau Fermi liquid property \cite{Baym:1975va,Schafer:2004jp},
\begin{eqnarray}
p_{iF} & = & (1-\kappa)\mu_{i}\,,\;\;\; i=u,d,\label{eq:pf-mu-qcd}\end{eqnarray}
where $\kappa=\frac{C_{F}\alpha_{S}}{2\pi}$, $\alpha_{S}$ is the
strong coupling constant and $C_{F}=(N_{c}^{2}-1)/(2N_{c})$ with
the number of colors $N_{c}=3$. The reduction in Fermi momentum means
a non-zero effective mass on the Fermi surface. This opens up phase
space characterized by the triangle inequality $p_{Fd}<p_{Fu}+p_{Fe}$.
As illustrated in Fig. \ref{cap:phasespace}, two dashed circles denote
the Fermi surfaces for free $d$ and $u$ quarks. They shrink to two
smaller solid circles after the interaction is turned on. The amount
of reduction in Fermi momentum for $d$ quarks is larger than that
for $u$ quarks, then there is a triangle among Fermi momenta implying
non-vanishing phase space. The corrections proportional to $\alpha_{s}$
is from the quark-quark forward scattering via one gluon exchange
with zero quark mass. 

In Ref. \cite{Wang:2006tg} we re-analyzed phase space for DU processes
by deriving $p_{F}$ as a function of $\mu$ in QCD with non-zero
quark masses and some other features not considered in Ref. \cite{Baym:1975va,Schafer:2004jp}.
We also carry out the same task in the NJL model. When $\alpha_{S}$
and $\kappa$ are small, $\kappa$ as a function $\mu$ can be solved,
\begin{eqnarray}
\kappa(\mu) & = & \left[\kappa(\mu_{0})-\frac{C_{F}\alpha_{S}}{2\pi}\right]\frac{\mu_{0}^{2}}{\mu^{2}}+\frac{C_{F}\alpha_{S}}{2\pi},\label{eq:kappa-qcd}\end{eqnarray}
One sees that if $\kappa(\mu_{0})<\frac{C_{F}\alpha_{S}}{2\pi}$ at
$\mu_{0}$, then $\kappa(\mu)<\frac{C_{F}\alpha_{S}}{2\pi}$ is always
true for any values of $\mu$. This is the physical branch, the other
one corresponds to $\kappa(\mu_{0})>\frac{C_{F}\alpha_{S}}{2\pi}$.
A similar result can also be found in NJL model, \begin{eqnarray}
\kappa(\mu) & = & \left[\kappa(\mu_{0})-\frac{4G_{S}\mu^{2}}{3\pi^{2}}\right]\frac{\mu_{0}^{2}}{\mu^{2}}+\frac{4G_{S}\mu^{2}}{3\pi^{2}},\label{eq:kappa0}\end{eqnarray}
where $G_{S}$ is the coupling constant for the scalar and pseudoscalar
channels in NJL model. As in the QCD case, $\kappa(\mu)<\frac{4G_{S}\mu^{2}}{3\pi^{2}}$
is the physical solution. Following Eq. (\ref{eq:kappa-qcd}) and
(\ref{eq:kappa0}), both physical solutions show that the Fermi momentum
reduction coefficient $\kappa$ are monotonously increasing functions
of the chemical potential. Such a trend implies that phase space for
neutrino emissions is quenched at lower baryon densities. The property
seems robust and independent of specific models in computing the Landau
coefficients. 

\section{Fermi model of weak interaction }

\label{sec:Fermi-four-fermion}
\begin{figure}
\caption{\label{cap:diagrams}The diagrams corresponds to the two terms in
Eq. (\ref{eq:4-fermion}). The dashed lines denote W-bosons. }
\includegraphics[scale=0.6]{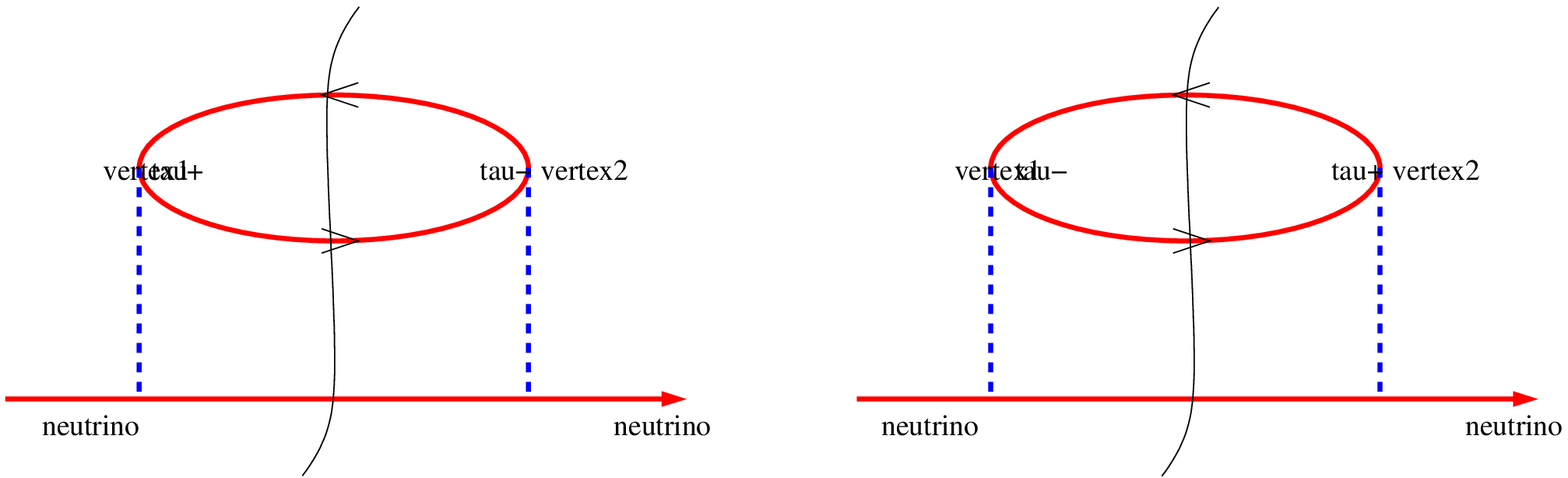}
\end{figure}

In calculating the neutrino emissivity
in Urca processes, it is convenient to use the Fermi effective model
of weak interaction since the characteristic energy scale is about
a few hundred MeV, much less than the W-boson mass. The interaction
Hamiltonian of the model is \begin{eqnarray*}
H_{I} & = & \frac{G}{\sqrt{2}}J^{\mu}J_{\mu}^{\dagger},\end{eqnarray*}
where the weak current is \begin{eqnarray}
J^{\mu}(x) & = & \overline{\psi}_{\nu}\gamma^{\mu}(1-\gamma_{5})\psi_{e}+\overline{\psi}_{u}\gamma^{\mu}(1-\gamma_{5})\psi_{d}\nonumber \\
 & = & \overline{\psi}_{l}\gamma^{\mu}(1-\gamma_{5})\tau_{+}\psi_{l}+\overline{\psi}_{q}\gamma^{\mu}(1-\gamma_{5})\tau_{+}\psi_{q},\nonumber \\
J_{\mu}^{\dagger}(x) & = & \overline{\psi}_{l}\gamma_{\mu}(1-\gamma_{5})\tau_{-}\psi_{l}+\overline{\psi}_{q}\gamma_{\mu}(1-\gamma_{5})\tau_{-}\psi_{q},\label{eq:jmu}\end{eqnarray}
where $\psi_{l}=(\psi_{\nu},\psi_{e})^{T}$ and $\psi_{q}=(\psi_{u},\psi_{d})^{T}$.
Here the flavor matrices $\tau_{\pm}$ are defined by \[
\tau_{-}=\left(\begin{array}{cc}
0 & 0\\
1 & 0\end{array}\right),\;\tau_{+}=\left(\begin{array}{cc}
0 & 1\\
0 & 0\end{array}\right).\]
The matrix elements for Urca processes amount to calculating 
\begin{eqnarray}
|M|^{2} & \rightarrow & \frac{G^{2}}{2}[J^{\mu}J_{\mu}^{\dagger}][J^{\nu}J_{\nu}^{\dagger}]\nonumber \\
 & \rightarrow & \frac{G^{2}}{2}\left[\overline{\psi}_{l}\gamma^{\mu}(1-\gamma_{5})\tau_{+}\psi_{l}\overline{\psi}_{l}\gamma_{\nu}(1-\gamma_{5})\tau_{-}\psi_{l}\overline{\psi}_{q}\gamma_{\mu}(1-\gamma_{5})\tau_{-}\psi_{q}\overline{\psi}_{q}\gamma^{\nu}(1-\gamma_{5})\tau_{+}\psi_{q}\right.\nonumber \\
 &  & +\overline{\psi}_{l}\gamma_{\mu}(1-\gamma_{5})\tau_{-}\psi_{l}\overline{\psi}_{l}\gamma^{\nu}(1-\gamma_{5})\tau_{+}\psi_{l}\nonumber\\
& & \left. \overline{\psi}_{q}\gamma^{\mu}(1-\gamma_{5})\tau_{+} \psi_{q} \overline{\psi}_{q}\gamma_{\nu}(1-\gamma_{5})\tau_{-}\psi_{q}\right],
\label{eq:4-fermion}
\end{eqnarray}
where we have only kept terms relevant to direct Urca. By renaming
indices $\nu\rightarrow\mu,\mu\rightarrow\nu$ in the second term,
one can prove that the second is identical to the first. These terms
correspond to two diagrams in Fig. \ref{cap:diagrams} in the quark
tensor, so we get a factor of 2, which we took into account in Eq.
(18) of Ref. \cite{Schmitt:2005wg}. 

\vspace{0.5cm}

\psfrag{tau+}{$\tau _+$}
\psfrag{tau-}{$\tau _-$}
\psfrag{vertex1}{$\mu$}
\psfrag{vertex2}{$\nu$}
\psfrag{neutrino}{$\nu _e$}

\section{Flavor sector of neutrino emissivity}

The flavor sector of neutrino emissivity in CSC is more complicated
than that in the normal phase because of pairings in flavor space.
Nambu-Gorkov (NG) basis is a convenient mathematical tool to describe
pairings. We have to write down the quark-quark-W-boson vertices in
NG basis. To this end, we enlarge the spinor space by introducing
a couplet field consisting of a quark field and its charge conjugate
partner, \begin{eqnarray*}
\Psi & = & \left(\begin{array}{c}
\psi\\
\psi_{c}\end{array}\right),\;\;\overline{\Psi}=(\overline{\psi},\overline{\psi}_{c}),\\
\psi_{c} & = & C\overline{\psi}^{T},\;\;\overline{\psi}_{c}=\psi^{T}C,\end{eqnarray*}
where $C=i\gamma^{2}\gamma^{0}$ is the charge conjugate operator.
Here we suppress the quark index $q$ in quark fields. Now we rewrite
the bilinear term $\overline{\psi}_{q}\gamma^{\mu}(1-\gamma_{5})\tau_{+}\psi_{q}$
in the weak current in Eq. (\ref{eq:jmu}), \begin{eqnarray*}
\overline{\psi}\Gamma_{+}^{\mu}\psi & = & \psi_{c}^{T}C\Gamma_{+}^{\mu}C\overline{\psi}_{c}^{T}=-\left[\overline{\psi}_{c}C^{T}\Gamma_{+}^{\mu T}C^{T}\psi_{c}\right]^{T}\\
 & = & -\overline{\psi}_{c}C^{T}\Gamma_{+}^{\mu T}C^{T}\psi_{c}\equiv\overline{\psi}_{c}\overline{\Gamma}_{+}^{\mu}\psi_{c},\end{eqnarray*}
where we defined $\Gamma_{+}^{\mu}=\gamma^{\mu}(1-\gamma_{5})\tau_{+}$
and $\overline{\Gamma}_{+}^{\mu}\equiv-C\Gamma_{+}^{\mu T}C=-\gamma^{\mu}(1+\gamma_{5})\tau_{-}$.
Therefore the weak vertex in NG basis is, \begin{eqnarray*}
\Gamma_{NG,\pm}^{\mu} & \rightarrow & \left(\begin{array}{cc}
\gamma^{\mu}(1-\gamma_{5})\tau_{\pm} & 0\\
0 & -\gamma^{\mu}(1+\gamma_{5})\tau_{\mp}\end{array}\right).\end{eqnarray*}

The polarization tensor {[}see, e.g. Eq. (12) of of Ref. \cite{Schmitt:2005wg}{]}
involved in computing the neutrino emissivity is written as 

\begin{eqnarray}
\Pi^{\mu\nu}(Q) & = & T\sum_{k_{0}}\int\frac{d^{3}\mathbf{k}}{(2\pi)^{3}}\mathrm{Tr}_{NG,c,f,s}[\Gamma_{NG,-}^{\mu}S(K)\Gamma_{NG,+}^{\nu}S(K+Q)]\nonumber \\
 & = & \Pi_{11}^{\mu\nu}+\Pi_{22}^{\mu\nu}+\Pi_{12}^{\mu\nu}+\Pi_{21}^{\mu\nu}\label{eq:polar}\end{eqnarray}
where the trace is over NG, color, flavor and Dirac indices. Note
that the factor $1/2$ arising from averaging in NG basis cancels
a factor of 2 from another identical term with a trace $\mathrm{Tr}[\Gamma_{NG,+}^{\mu}S(K)\Gamma_{NG,-}^{\nu}S(K+Q)]$,
see Eq. (18) of Ref. \cite{Schmitt:2005wg} and the argument that
follows. $\Pi_{11}^{\mu\nu}$ and $\Pi_{22}^{\mu\nu}$ are diagonal
components of the polarization tensor which are there in the normal
phase, while $\Pi_{12}^{\mu\nu}$ and $\Pi_{21}^{\mu\nu}$ are off-diagonal
components proportional to condensate square and vanishing in the
normal phase. As is shown in Tab. \ref{cap:flavor-trace} for the
flavor trace, one can easily verify that for single flavor or spin-one
pairings \cite{Pisarski:1999tv,Schafer:2000tw,Alford:2002rz,Schmitt:2002sc,Buballa:2002wy,Schmitt:2003xq,Schmitt:2004et,Aguilera:2005tg,Alford:2005yy}
off-diagonal parts are absent, but it is not the case for spin-zero
pairings such as the color-flavor-locked phase {[}see, e.g. Ref. \cite{Alford:1998mk}{]}
or the 2-flavor CSC phase {[}see, e.g. Ref. \cite{Pisarski:1999tv,Wang:2001aq}{]}.
The reason is the electric charge conservation \cite{Schmitt:2005wg}. 

\begin{table}
\caption{\label{cap:flavor-trace}Flavor traces in neutrino emissivity in
color superconducting phases. The order parameter is denoted by $\Delta$.
$\mathbf{I}$ are $SU(3)_{c}$ generators in color space with $(I_{i})_{jk}=-i\epsilon_{ijk}$
and $\mathbf{J}$ are $SU(3)_{f}$ generators in flavor space with
$(J_{i})_{jk}=-i\epsilon_{ijk}$. }
\begin{tabular}{|c|c|c|c|}
\hline 
& $\Delta$& $\mathrm{Tr}_{f}[\tau_{\pm}\Delta\tau_{\pm}\Delta^{\dagger}]$&
off-diagonal\tabularnewline
\hline 
CFL& $\mathbf{J}\cdot\mathbf{I}$& 2& $\surd$\tabularnewline
\hline 
Single Flavor& $\begin{array}{c}
\delta_{fu}\delta_{gu}\\
\delta_{fd}\delta_{gd}\end{array}$& 0& $\times$\tabularnewline
\hline
\end{tabular}
\end{table}

{\bf Acknowledgements. } 
The author thanks Shi Pu for reading the manuscript and helpful discussions. 
The author is supported in part by the startup grant from 
University of Science and Technology of China (USTC) in 
association with 'Bai Ren' project of Chinese Academy of Sciences (CAS).

\end{document}